%A comment on matrix strings
%by Thomas Wynter
\input harvmac
\input epsf
%\draftmode

\def\R{\relax{\rm I\kern-.18em R}}
\font\cmss=cmss10 \font\cmsss=cmss10 at 7pt
\def\Z{\relax\ifmmode\mathchoice
{\hbox{\cmss Z\kern-.4em Z}}{\hbox{\cmss Z\kern-.4em Z}}
{\lower.9pt\hbox{\cmsss Z\kern-.4em Z}}
{\lower1.2pt\hbox{\cmsss Z\kern-.4em Z}}\else{\cmss Z\kern-.4em
Z}\fi}\
\def\np{Nucl. Phys. }
\def\pl{Phys. Lett. }
\def\pr{Phys. Rev. }

\def\prl{Phys. Rev. Lett. }

\def\Tr{{\rm Tr}}

\def\CN{{\cal N}}

\def\CN0{{\cal N}_0}

\rightline{hep-th/9709029}

\Title{}
{\vbox{
\centerline{Gauge fields and interactions in matrix string theory}
}}
%%%%%%%%%%%%%%%%%%%%%%%%%%%%%%%%%%%%%%%%%%%%%%%%%%%%%%%%
\lref\witt{E.~Witten, \np {\bf B443}(1995) 85-126, hep-th/9503124.}
\lref\bsfs{T.~Banks, W.~Fischler, S.H.~Shenker, and L.~Susskind, \pr
{\bf D55}(1996) 5112-5128, hep-th/9610043.}
\lref\pol{J.~Polchinski, \prl {\bf 75}(1995) 4724-4725, hep-th/9510017.}
\lref\wittd{E.~Witten, \np {\bf B460}(1996) 335-350, hep-th/9510135.}
\lref\DFSKPDS{U.H.~Danielsson, G.~Ferretti and B.~Sudborg, Int. J. Mod. 
Phys. {\bf A11}(1996) 5463-5478, hep-th/9603081; D.~Kabat and 
P.~Pouliot, \prl {\bf 77}(1996) 1004-1007, hep-th/9603127; M.R.~Douglas,
D.~Kabat, P.~Pouliot and S.H.~Shenker, \np {\bf B485}(1997) 85-127,
hep-th/9608024.}
\lref\motl{L.~Motl, hep-th/9701025.}
\lref\bs{T.~Banks, and N.~Seiberg, \np {\bf B497}(1997) 41-55,
hep-th/9702187.} 
\lref\dvsq{R.~Dijkgraaf, E.~Verlinde and H.~Verlinde, hep-th/9703030.}
\lref\wati{W.~Taylor, \pl {\bf B394}(1997) 283-287, hep-th/9611042.}
\lref\mrd{M.R.~Douglas, hep-th/9707228.}
\lref\thooft{G.~'t Hooft, \np {\bf B190}(1981) 455-478.}
\lref\mand{S.~Mandelstam, Prog. of Theor. Phys. Suppl. No. 86 (1986) 
163-170.}
\lref\gsw{M.B.~Green, J.H.~Schwarz, E.~Witten, Superstring Theory, 
Cambridge University Press (1987).}
\lref\kni{V.G.~Knizhnik, Sov. Phys. Usp. {\bf 32}(11)(1989) 945-971.}

\vskip6pt

\centerline{Thomas Wynter\footnote{$^\circ$}{{\tt
wynter@wasa.saclay.cea.fr}}\footnote{$^{\dagger}$}{This work is 
partially support by the E.E.C. contract TMR ERBFMRXCT960012}
}

\centerline{{\it  Service de Physique Th\'eorique, C.E.A. - Saclay,
  F-91191 Gif-Sur-Yvette, France}} 

\vskip .3in
\baselineskip10pt{ It is
shown that all possible N sheeted coverings of the cylinder are
contained in type IIA matrix string theory as non-trivial gauge field
configurations. Using these gauge field configurations as backgrounds
the large $N$ limit is shown to lead to the type IIA conformal field
theory defined on the corresponding Riemann surfaces. The sum over
string diagrams is identified as the sum over non-trivial gauge
backgrounds of the SYM theory. }
\bigskip
\rightline{SPhT-97/105}

%\draft
\Date{August 1997 }

\baselineskip=16pt plus 2pt minus 2pt
\bigskip

%%%%%%%%%%%%%%%%%%%%%%%%%%%%%%%%%%%%%%%%%%%%%%%%%%%%%%
\newsec{Introduction}

Matrix theory\bsfs\ is a concrete proposal for a non-perturbative
description of M theory\witt , the 11 dimensional theory postulated as
the strong coupling limit of type IIA string theory. Formulated in the
infinite momentum frame it describes M-theory by the
$N\rightarrow\infty$ limit of the supersymmetric quantum mechanics of
D0 branes\pol\wittd\DFSKPDS .

Type IIA string theory is recovered from M theory by compactifying
M theory on a circle and identifying the string coupling constant
$g_s$ with the radius of compactification. One is thus naturally led
to study matrix theory compactified on a circle \motl\bs\dvsq . The 2d
SYM theory resulting from this compactification should give a
non-perturbative description of type IIA string theory. This was
studied in the most depth in the elegant paper by Dijkgraaf, Verlinde
and Verlinde \dvsq . The string coupling constant in this model is
related to the the SYM coupling by $g_s=1/\sqrt{\alpha'}g_{QCD}$.  By
looking at the strong coupling limit of the SYM they argued that
perturbative string theory could indeed be recovered. The deep
infra-red (strong coupling) limit of the theory was identified as an
$S_N$ orbifold conformal theory describing freely propagating
strings. From this infra-red limit the original SYM theory is seen as
an irrelevant perturbation of the conformal field theory. They further
argued that this irrelevant perturbation corresponded to the splitting
and joining of strings and constructed a local operator to describe
such an interaction. By dimensional analysis they showed that it
would, as required, be associated with a linear power of the string
coupling constant $g_s$.

In this paper we aim to understand directly from the SYM theory how
string interactions emerge. We start in section 2 by recalling the
model. In section 3 we study the classical solutions of the theory and
identify not only the solutions corresponding to multiply wound (long)
strings but also those corresponding to arbitrary string
interactions. They are characterized by particular non-trivial gauge
field configurations which we show how construct explicitly. In
section 4 we turn to the effective action for the low energy
excitations of the theory in the limit $N\rightarrow\infty$. We show
that they are described by the orbifold conformal field theory defined
on the multisheeted Riemann surface specified by the background gauge
field configuration, irrespective of the value of $g_s$. The sum over
string diagrams is identified as the sum over non-trivial gauge
configurations of the SYM theory.

\newsec{$N=8$ two dimensional gauge theory as non-perturbative string
theory}
In this section we briefly summarize the principle ideas of matrix 
string theory as first proposed in \motl\ and \bs\  and  
developed most fully in \dvsq . The starting point is 10-d
supersymmetric Yang-Mills theory dimensionally reduced to two
dimensions: 
\eqn\matrixs{
S=\int\,d\tau d\sigma
\Tr\bigl[ {1\over 2}(D_{\alpha}X^I)^2+{i\over 2}\Theta^T\slash\hskip -7pt
D\Theta-{1\over 4}F_{\alpha\beta}^2+{1\over 4 g_s^2}[X^I,X^J]^2
+{1\over 2g_s}\Theta^T\gamma_i[X^I,\Theta]\bigr].
}
This can be obtained by compactifying matrix theory on a circle (see
\bsfs\wati ). All fields are $N\times N$
hermitean matrices. The index $I$ runs from $1$ to $8$ and the 16
component fermion fields split into the $8_s$ and $8_c$
representations of $SO(8)$.  $g_s$ is the string coupling constant and
the coordinate $\sigma$ runs from $0$ to $2\pi$.

Weakly coupled strings are to be recovered by considering the limit
$g_s\rightarrow 0$ or equivalently the infra-red limit of the
theory. It was argued in the above papers that the limiting point
$g_s=0$ would be described by a conformal theory in which all matrices
commute.  The matrix coordinates can then be written as:
\eqn\XUxU{
X^I=Ux^IU^{\dagger}\quad{\rm and}\quad\Theta=U\theta U^{\dagger},
}
where $x^I$ and $\theta$ are diagonal matrices and $U$ is a unitary
matrix. The abelian remnant of the gauge fields would decouple leaving
$N$ copies of the light cone Green Schwarz action. We will discuss
this reduction in more detail in section 4.

As was first realised in \motl\ it is possible to find field
configurations \XUxU\ corresponding to strings wound multiply around the
compact direction $\sigma$. Specifically we can choose the matrix $U$
in \XUxU\ such that
\eqn\UGamU{
U(\sigma+2\pi)=U(\sigma) g\quad\Rightarrow\quad x(\sigma+2\pi)
=g x(\sigma) g^{\dagger}
}
where $g$ is an element of the Weyl group of $U(N)$. The result being
that in going around the compact direction the eigenvalues are
interchanged and thus form cycles of varying length. For a cycle of
length $n$ consisting of the eigenvalues $x_1,\cdots,x_n$ one has 
\eqn\xcyc{
x_i(\sigma+2\pi)=x_{i+1}(\sigma)\quad{\rm with}\quad x_{n+1}=x_1.
}

In the infra red limit we thus would expect a conformal field theory
whose Hilbert space decomposes into ``twisted sectors'' given by the
different ways of partitioning the $N$ eigenvalues into cycles:
\eqn\NnNn{
N=\sum_N n N_n,}
with $N_n$ the number of cycles of length n. For each of these twisted
sectors the original non-abelian gauge symmetry is broken down to a
discrete $S_N$ symmetry which acts by (a) permuting cycles of
identical length, and by (b) cycling through the eigenvalues in a
given cycle, i.e. $x_i\rightarrow x_{i+1},\quad i=1,\cdots,n\,\,$ for the
cycle described above.

The outcome is that the conformal field theory in the infra-red
$(g_s=0)$ limit is identified as the $N=8$ supersymmetric sigma model
defined on the orbifold target space
\eqn\Sorb{
S^N{\bf R}^8=({\bf R}^8)^N/S_N.}

The final step in identifying freely propagating strings is to send
$N\rightarrow\infty$ and to keep only the lowest energy excitations
i.e. those corresponding to strings of length ${\cal O}(N)$. The
length of an individual string divided by the total length of all the
strings is then identified with the light cone $p^+$ momenta and the
discrete remnant gauge symmetry (b), described above, becomes the
symmetry under translation $\sigma\rightarrow\sigma + const$ of the
string world sheet.

String interactions arise when two eigenvalues
approach each other and interchange. This is illustrated in Fig. 1
where for clarity we have chosen $N=2$. 
In this case the interchange
of eigenvalues describes the transition between an incoming state
consisting of a single string and an outgoing state consisting of two
strings. At the point where the eigenvalues meet (marked by the cross
in the diagram) a $U(2)$ subgroup of the otherwise completely broken
$U(N)$ symmetry is restored.  The
authors of \dvsq\ argued that near the infra-red
fixed point (where the non-abelian part of the SYM theory is partially
restored) the theory would be described by the conformal field
theory perturbed by an irrelevant operator corresponding to such
an interaction. Using twist operators for the bosonic and fermionic
fields they constructed a Lorentz invariant supersymmetric operator
and showed that it was the unique least irrelevant operator of this type.
Furthermore its dimension is such that its
contribution to the action is naturally associated with a linear
power of the string coupling constant $g_s$.

\newsec{Gauge field configurations and string interactions}

The objective of this paper is to understand directly from
the SYM theory how string interactions arise. We begin by 
investigating the non-trivial gauge configurations of the theory. We
will identify them by studying the classical equations of motion of
\matrixs . In particular we will construct field configurations
corresponding to type IIA string theory defined on arbitrary $N$
sheeted Riemann surfaces, where $N$ is the size of the matrix fields.

We restrict our attention to field configurations in which
all bosonic and fermionic matrices commute.

Let us start with the free strings wrapped multiply around the $\sigma$
direction \UGamU\ and focus on one particular block corresponding to a
single cycle of length $n$. It is trivial to find such 
configurations which also satisfy the classical equations of
motion. We will do
this just for the bosonic fields. The generalization to the fermions
is obvious.
We start from the configuration 
\eqn\freesol{
A_{\alpha}=0
\quad{\rm and}\quad X(\sigma,\tau)={\rm diag}(x_1(\sigma,\tau),
\cdots ,x_n(\sigma,\tau))
\quad{\rm with}\quad x_i(\sigma+2\pi,\tau)=x_{i+1}(\sigma,\tau).}
which is multivalued and with $X$ satisfying the equations of motion,
$\partial_{\alpha}\partial^{\alpha}X=0$. Using the matrix $U(\sigma)$,
which satisfies 
\eqn\defU{
U(\sigma+2\pi)=U(\sigma)g\quad{\rm with}\quad
g=\pmatrix{0&1&0&\ldots&0\cr
           0&0&1&\ldots&0\cr
            \vdots&\ddots&\ddots&\ddots&\vdots\cr
            1&0&\ldots&0&0\cr},}
we gauge transform \freesol\ to arrive at the singlevalued
configuration
\eqn\longs{
X=U(\sigma){\rm diag}(x_1(\sigma,\tau),\cdots ,x_n(\sigma,\tau))
U^{\dagger}(\sigma),
\quad
A_{\sigma}=igU^{\dagger}(\partial_{\sigma}U)
\quad{\rm and}\quad
A_{\tau}=0,}
which satisfies the non-abelian equations of motion $D_{\alpha}
D^{\alpha}X=0$.  We see that multiply wound strings are associated
with a non-zero pure gauge field $A_{\alpha}$ \longs\ which cannot be
gauge transformed away by a single valued gauge transformation.

For convenience we now make the standard change of
coordinates from cylindrical coordinates $\sigma$, $\tau$ to $z$ and
$\bar{z}$ defined in the complex plane by $z=e^{\tau+i\sigma}$. 
The $N$ eigenvalues will then form an $N$ sheeted covering of
the complex plane. The winding sectors are characterized by having
branch points of different orders which are placed at the origin and
which connect together the sheets into cycles. The order of the branch
cut determines the length of the corresponding cycle. 

In this picture string interactions are trivial, they would correspond
to having branch cuts occurring away from the origin (see fig. 1). 
What is not
immediately obvious, however, is whether there exist singlevalued
matrix configurations whose multivalued eigenvalues correspond to such
a Riemann surface.
\vskip 40pt
\hskip 20pt
\epsfbox{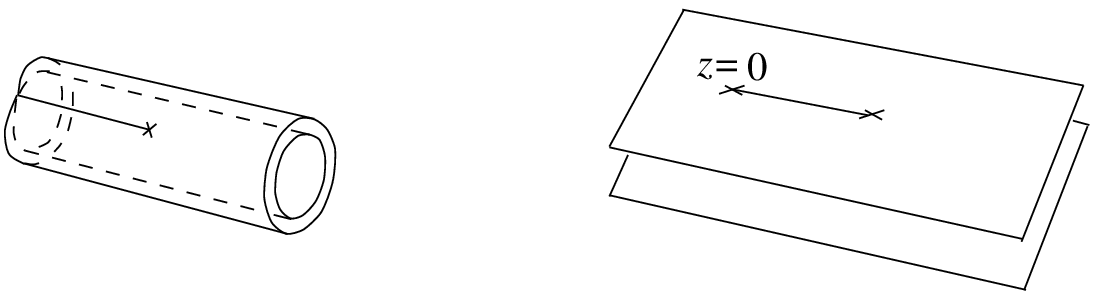}
\vskip 5pt
\centerline{Fig. 1. (a) A simple string interaction and (b) its
representation in the complex plane}
\vskip 10pt

Below we demonstrate that this is indeed the case. We will show that
any $N$ sheeted Riemann surface connected together by an arbitrary set of
branch points can be described by the multivalued eigenvalues of a
single-valued hermitean matrix. We will achieve this by
construction. Specifically we will search for single-valued matrices,
$X$, whose eigenvalues are multivalued holomorphic functions of $z$
and which furthermore are hermitean in the sense that
$X(z,\bar{z})^{\dagger}=X(\bar{z},z)$.

The starting point is the fact that an arbitrary $N$ sheeted Riemann
surface can always be generated by a polynomial equation of degree $N$
whose coefficients are single-valued functions of $z$.  We thus search
for single valued matrix solutions of the equation
\eqn\poly{
\sum_{n=0}^N a_n X^n=0,
}
with $a_n(z)$ single valued and $a_N=1$. If such a solution, $X$, 
exists its $N$ eigenvalues
$x_j$, $(j=1,\cdots ,N)$ will
correspond to the  $N$ solutions of \poly\ and together 
will describe the $N$ sheeted Riemann surface. Furthermore since they are
holomorphic functions of $z$ they will trivialy satisfy the equations
of motion: $\partial\bar{\partial}x(z)=0$. Thus, in complete analogy
with the winding sectors discussed above, the non-abelian equations of
motion will be satisfied by the field configuration 
\eqn\Xsing{
X=U\,{\rm diag}(x_1,\cdots ,x_N)U^{\dagger},
\quad A_{\alpha}=igU^{\dagger}(\partial_{\alpha}U).}

We now demonstrate that such a solution exists. We start by solving
\poly\ for complex matrices $M$. It is trivial to see that the matrix
\eqn\Mdef{
M=\pmatrix{-a_{N-1}&-a_{N-2}&\ldots&-a_1&-a_0\cr
            1&0&\ldots&0&0\cr
           0&1&\cdots&0&0\cr
           \vdots& &\ddots&\vdots&\vdots\cr
           0&0&\ldots&1&0\cr}}
satisfies equation \poly\footnote{$^3$}{I thank Michel Bauer for
suggesting this matrix}. The matrix $M$ can be diagonalized by the
complex invertible matrix $S$:
\eqn\Sdef{
M=S\,{\rm diag}(x_1,\cdots ,x_N) S^{-1}\quad{\rm with}\quad
S=\pmatrix{x_1^{N-1}&x_2^{N-1}&\ldots&x_N^{N-1}\cr
           x_1^{N-2}&x_2^{N-2}&\ldots&x_N^{N-2}\cr
           \vdots&\vdots&\ddots&\vdots\cr
           1&1&\ldots&1}}
where the $x_i$, $i=1,\cdots,N$ are the $N$ eigenvalues of the matrix
$M$, i.e. the $N$ scalar solutions of the polynomial equation \poly . 
Note that strictly this matrix is only invertible away from the
branch points where two or more eigenvalues coincide.  Now let us
examine what happens about the branch points of the Riemann
surface. In circling a branch point some of the eigenvalues, $x_i$ are
interchanged
\eqn\ssg{
{\rm diag}(x_1,\cdots ,x_N)\rightarrow g\,{\rm diag}(x_1,\cdots
,x_N)\,g^{\dagger}\quad{\rm and}\quad S\rightarrow S g,}
where $g$ is, as before, an element of the Weyl group of $U(N)$. We now
decompose $S$ into a hermitean matrix (positive definite) times a
unitary matrix
\eqn\shu{
S(z)=H(z,\bar{z})U(z,\bar{z}). }
Since this is a unique decomposition it follows from \ssg\ (and the
fact that $g$ is an element of $U(N)$) that in circling a branch point
\eqn\hug{
H(z,\bar{z})\rightarrow H(z,\bar{z})\quad{\rm and}\quad
U(z,\bar{z})\rightarrow U(z,\bar{z}) g. }
From the matrix $U(z,\bar{z})$ \shu\ and the matrix ${\rm
diag}(x_1(z),\cdots ,x_N(z))$ we then construct the single valued
solution \Xsing . We illustrate this construction in the appendix with
a simple example, that of $Z_N$ Riemann surfaces.

The most important result of this construction is not the
particular matrix $X$ \Xsing\ but rather the matrix $U$ \shu .  We
have shown that for every $N$ sheeted Riemann surface there exists a
unitary matrix $U$ which generates the correct monodromies around the
branch points. Thus an arbitrary function defined on an $N$ sheeted
Riemann surface, covering the complex plane, can always be described
by a single valued $N\times N$ matrix.  Furthermore each Riemann
surface is characterized by a particular (up to gauge transformations)
non-trivial gauge field configuration
$A_{\alpha}=igU^{\dagger}(\partial_{\alpha}U)$, which interpolates
between the winding sector in the infinite past $(z=0)$ and the
winding sector in the infinite future $(z=\infty)$. 
These field configurations are singular with the gauge fields
diverging at the branch points. Since these are
non-trivial gauge configurations they will necessarily play a crucial
role in the effective action for the low energy excitations.

\newsec{The effective action}

The key idea is that string scattering amplitudes are calculated as
SYM correlation functions connecting together different winding
sectors. The sum over all the possible string interactions is
reproduced by the sum over all non-trivial gauge configurations
connecting the winding sector in the infinite past with the winding
sector in the infinite future. 

Let us therefore study the effective action calculated about one of
the non-trivial gauge field configurations of the previous section. We
focus on a Riemann surface $\Sigma$ corresponding to a few strings
wound multiply around the $\sigma$ direction which interact at one or
two points.  Denoting by $U_{\Sigma}$ its associated gauge field
configuration we define field variables, $X_I^{\Sigma}$,
$\Theta^{\Sigma}$ and $A_{\alpha}^{\Sigma}$ for which the
structure of the Riemann surface is manifest:
\eqn\gbgrd{\eqalign{
X_I=&U_{\Sigma}X_I^{\Sigma}U_{\Sigma}^{\dagger}\cr
\Theta=&U_{\Sigma}\Theta^{\Sigma}U_{\Sigma}^{\dagger}\cr
A_{\alpha}=&U_{\Sigma} A_{\alpha}^{\Sigma}U_{\Sigma}^{\dagger}
     -ig_s(\partial_{\alpha}U_{\Sigma})U_{\Sigma}^{\dagger}.}}
These are well defined 
everywhere except at the interaction points, where the transformation
is singular.
Their diagonal elements, by definition,
live on the $N$ sheeted Riemann surface $\Sigma$. Their off-diagonal
elements connect together the different sheets(diagonal elements). We
illustrate this in Fig. 2 where for clarity we have unwound the
multiply wound long strings and have illustrated the part of
the resulting Riemann surface surrounding an interaction point. The
vertical wall represents a square root cut.
\vskip 40pt
\hskip 20pt
\epsfbox{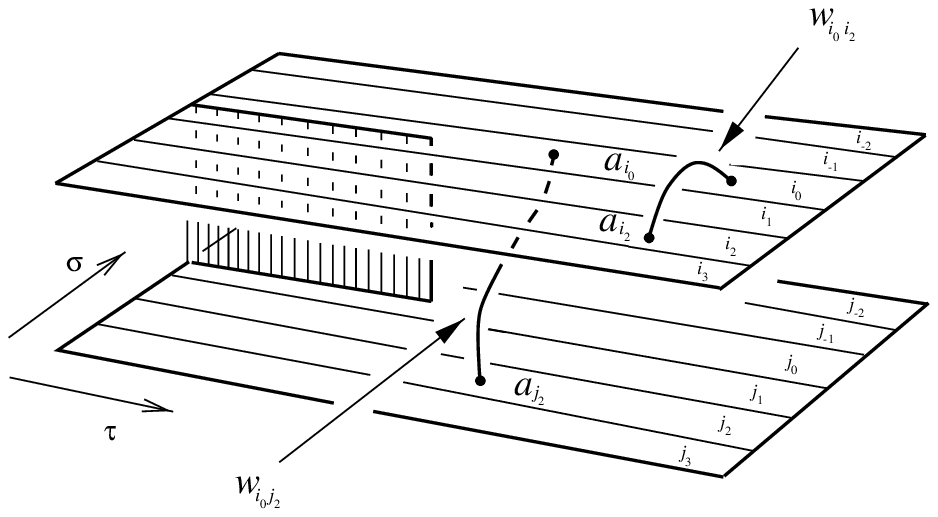}
\vskip 5pt
\centerline{Fig. 2 Diagonal $(a_i)$ and off diagonal $(w_{ij})$ elements
defined on the Riemann surface $\Sigma$.}
\vskip 20pt
The strips are of width $2\pi$ in the $\sigma$ direction with each
strip labeled by an index $i$ or $j$ specifying the diagonal element
$a_i$ defined on it. The ``strings''\footnote{$^1$}{The action
\matrixs\ can be thought of as the action for D-strings. The
``strings'' are then the fundamental type IIB strings stretched
between the D-strings} connecting together the different strips
represent the off diagonal elements $w_{ij}$ with indices given by the
strips on which they end. In terms of the Riemann surface in which the
long strings have been unwound they are bilocal fields and can connect
together a long string to itself or to another long string.

Ultimately we are interested in the $N\rightarrow\infty$ limit where
non-compact ten-dimensional space is recovered. The limit is taken
while rescaling the coordinates $(\sigma,\tau)$ by $1/N$ so that only
states invariant under change of $N$ are left. This also allows one to
equate the string length with the $p^+$ momentum carried by the string
as required by light cone string theory. The large $N$ limit is thus
the infra-red fixed point of the theory with invariance under change of
$N$ equivalent to conformal invariance\footnote{$^2$}
{See \mrd\ where sigma model versions
of matrix theory are studied and the Ricci flatness condition is
recovered as a fixed point of the large $N$ renormalization group of
the theory}. There is of course a subtlety in that the dimension of
the matrix fields is increasing as $N$ and could compensate for the
renormalization group flow.

In the light of the above comment let us now turn to the determination
of the effective action paying particular attention to the large $N$
limit. Roughly speaking we would like to integrate out the
off-diagonal elements of the matrices to arrive at an effective action
for the diagonal elements. This will only in fact be possible for
configurations in which the diagonal elements do not touch.

For compactness we write the action \matrixs\ in it's ten-dimensional
form (we drop from here on the superscript $\Sigma$):
\eqn\tenym{
S=\int\,d^2\sigma \Tr\bigl[-{1\over 4}F_{\mu\nu}^2+{i\over 2}
\Theta^T\slash\hskip -7pt D \Theta\bigr],
}
where $\mu,\nu=0,\cdots,9$. The indices split into $\alpha,\beta=0,9$
and $I,J=1\cdots 8$. The fields depend upon the coordinates
$\sigma_{\alpha}=(\tau,\sigma)$ and the gauge fields $A_I$ are the
bosonic coordinates $X_I$ of \matrixs .
We then decompose the fields into their diagonal
components ($a$ and $\theta$) and their off diagonal components ($w$
and $\eta$) :
\eqn\Aw{\eqalign{
A_{\mu}=&a_{\mu}+ w_{\mu},\cr
\Theta=&\theta+ \eta,}}
and choose an abelian gauge\thooft\ to privilege the diagonal
elements. Specifically we add the gauge fixing term
\eqn\gf{
{1\over 2}(\partial_{\mu}a^{\mu})^2+(D_{\mu}w_{\mu})^2
\quad{\rm where}\quad D_{\mu}=\partial_{\mu}-{i\over
g}[a_{\mu},\,\,\,\,\,]
}
We then write the action as an integral over the Riemann surface
$\Sigma$ and replace, for $N$ large, the indices of the off diagonal
elements by continuous coordinates corresponding to having a ``string''
connecting the arbitrary points $\sigma$ and $\sigma'$ of $\Sigma$ (at
equal $\tau$). The sum over indices is then replaced by an integral
over all such positions with coordinates rescaled by $1/N$. 
For example
\eqn\wscal{
\int\,d\tau d\sigma\Tr\bigl[{1\over g^2}[a_{\mu},w_{\nu}]^2\bigr]
\rightarrow
N^3\int\,d\tau d\sigma d\sigma'
{1\over g^2}(a_{\mu}(\sigma,\tau)-a_{\mu}(\sigma',\tau))^2
|w_{\nu}(\sigma,\sigma',\tau)|^2,}
where the factor of $N^3$ comes from rescaling the three coordinates.
Focusing on the purely bosonic part of the action we find that for
non-coinciding $a^{\mu}$ there is a single term (the above mass term)
in the integral over $w^{\mu}$ that dominates all the others. This is
made clear by rescaling the $w^{\mu}$ field by $1/ N^{3/2}$ (see
appendix). The bosonic part of the action then takes the form
\eqn\Sb{
S_{\rm b}=\int\bigl[-{1\over 2}(\partial_{\mu}a_{\nu})^2
                     -{1\over 2}(\nabla_{\mu}w_{\nu})^2
                      +{\cal O}\bigl({1\over\sqrt{N}}\bigr)\bigr],
\quad{\rm where}\quad
\nabla_{\mu}={1\over N}\partial_{\mu}-{i\over g}[a_{\mu},\,\,\,\,\,].
}
For the complete action (see Appendix) one finds
\eqn\Strexp{
S=S_{\rm G.S.}+S_{\rm g.f.}+S_0+{\cal O}\bigl({1\over\sqrt{N}}\bigr),}
where $S_{\rm G.S.}$ is the Green Schwarz action defined on the Riemann
surface $\Sigma$, $S_{\rm g.f.}$ is the gauge fixed action for the
abelian gauge fields which decouple and $S_0$ is given by
\eqn\Sgsgo{
S_0=\int\,d^2\sigma\,\Tr\bigl[-{1\over 2}(\nabla_{\mu}w_{\nu})^2
                              -{i\over 2}\eta\gamma_{\mu}\nabla^{\mu}
                              +\omega^*(\nabla_{\mu})^2\omega\bigr]}
with $\omega$ off diagonal ghost fields from the Fadeev-Popov
determinant. Providing 
$|a_i^{\mu}-a_j^{\mu}|^2\neq 0\quad\forall i\neq j$
we can drop the ${\cal O}({1\over\sqrt{N}})$ terms in \Strexp\ and the
integrals over $w^{\mu}$, $\eta$ and $\omega$ can be performed with
the determinants canceling due to supersymmetry in the usual way. 

The conclusion is that away from the interaction points, where
diagonal elements meet, the effective action is the Green Schwarz
action defined on the Riemann surface $\Sigma$, irrespective of the
value of $g_s$. 

At this level there is no way of determining what weight to associate
with a given Riemann surface and it might even appear that Riemann
surfaces for freely propagating strings should be treated on an equal
footing with those corresponding to interactions. What distinguishes
them of course is the singularities. In light cone string theory the
amplitudes have to be carefully calculated around the interaction
points, a cut off length $\epsilon$ being introduced and an operator
inserted at the interaction point to preserve supersymmetry and
Lorentz invariance\mand\gsw . From the point of view of the SYM theory
we know that the Riemann surface interpretation breaks down around the
branch point and thus we can expect the SYM to provide a
regularization of the singularity. To explicitly calculate this from
the SYM theory would involve including all the terms of ${\cal
O}(1/\sqrt{N})$ of
\Strexp\ (see the appendix). It is not yet known how to do this. 
In \dvsq\ however it was argued on dimensional grounds that
interactions would be associated with a factor of $g_s$, which is the
only constant in the SYM and which has the dimension of length.  This
was achieved by constructing a supersymmetric twist operator
associated with the branch point and
demonstrating that it has the correct dimension to appear with a
factor of $g_s$. What is not clear from this line of reasoning however
is what power of $N$ should be associated with the string coupling
constant. Indeed the large $N$ limit has been argued above to
correspond to the infra-red fixed point. The string interaction
operator (including the integrals over its position), has dimension 1,
and thus might simply be scaled away with a factor of $1/N$.

In terms of the unwound long strings of Fig. 2 the interaction weight
comes from integrating out the off diagonal elements (``strings'') in
a small region around the branch point. As well as the ``strings''
that connect the two sheets at points directly over each other there
are also the ``strings'' that connect points shifted with respect to
each other by an integer number of strips. As $N\rightarrow\infty$ the
density of such ``strings'' that contribute around the interaction
point increases with a factor of $N$. This could lead to a canceling
of the factor of $1/N$ coming from scaling the coordinates. A further
factor that could be important is the measure associated with the
integration over the position of the branch points. It should be
determined by restricting the gauge field integration to the field
configurations associated with a string interaction.  Clearly more
work needs to be done in these directions to resolve this question
satisfactorily.

\newsec{Conclusions}
We have set out to understand in more detail how string theory emerges
from matrix theory compactified on a circle.  The principle result of
this paper is that all the possible Riemann surfaces associated with
light cone string interactions are contained in the SYM as non-trivial
gauge field configurations. Treating these configurations as
backgrounds we have demonstrated that in the large $N$ limit the
effective action is described by the conformal field theory defined on
the associated Riemann surface. The weight to be associated with a
branch point can be deduced, up to its dependence on $N$, on
dimensional grounds by following the reasoning in \dvsq . The crucial
issue of what power of $N$, if any, is associated with the branch
points remains to be addressed.

Finally it is interesting to note that matrix description of string
theory appears to be a concrete realization of a proposal for 
the sum over all string diagrams suggested ten years ago by Knizhnik\kni\
\footnote{$^4$}{I thank Ivan Kostov for drawing my attention to
this paper}. He considered the $N\rightarrow\infty$ limit of string theory
defined on an $N$ sheeted covering of the complex plane and speculated
on whether it would contain non-perturbative phenomena.

\newsec{Acknowledgements} 
It is a great pleasure to thank Denis Bernard, Michel Bauer, and Ivan
Kostov for many helpful discussions. I would also like to thank Mike
Douglas for a discussion on the large $N$ limit. Finally I would like
to thank the CERN theory division for their hospitality, where this
work was begun.

\newsec{Appendix}

\subsec{$Z_N$ Riemann surfaces}
We illustrate the construction of section 3 on the simplest subset of
general $N$ sheeted Riemann surfaces, those with a $Z_N$ symmetry,
i.e. Riemann surfaces where all branch cuts are of order $N$.
In terms of the polynomial equation \poly\ they are described by the
single valued matrix solution to the equation
\eqn\ZNpoly{
X^N=a(z),}
with the position of the branch points given by the zeros of
$a(z)$. Defining the coordinate $\phi$ as the argument of $a(z)$,
\eqn\arga{
a(z)=|a(z)|e^{i\phi},}
we have for the matrix $S$ \Sdef\ 
\eqn\ZNS{
S_{kl}=|a|^{(1-{k\over N})}e^{i(1-{k\over N})\phi}\omega^{kl},}
where $\omega$ is the $N$th root of unity. Using equations \shu\ 
and \Xsing\ we find
\eqn\ZNU{
U_{kl}={1\over\sqrt{N}}e^{i(1-{k\over N})\phi}\omega^{kl}
\quad{\rm and}\quad
\bigl(A_{\phi}\bigr)_{kl}
=\cases{-{g\over 2}\bigl(1-{1\over N}\bigr),& 
for $k=l$;\cr
{g\over 2N}\bigl(1-i\cot\bigl[{\pi\over N}(k-l)\bigr]\bigr),& 
for $k\neq l$.\cr}}
For the special case of a winding sector consisting of several cycles
the unitary matrix $U$ is composed the sub-blocks for the
individual cycles placed down the diagonal. Each sub-block is given by
\ZNU\ with $N$ the length of the cycle and $\phi$ the original
cylindrical coordinate $\sigma$.

More general cases involve solving polynomial equations of degree 3 or
higher.

\subsec{The gauge fixed Lagrangian}
The total Lagrangian is given by
\eqn\Saw{
S=S_{\rm b}+S_{\rm f}+S_{\rm gh},}
where the bosonic and fermionic parts are given by
\eqn\SbN{\eqalign{
S_{\rm b}=&
\int\,\Tr\bigl[-{1\over 2}(\partial_{\mu}a_{\nu})^2
                      -{1\over 2}(D_{\mu}w_{\nu})^2
                    +{i\over g}[w_{\mu},w_{\nu}]\partial^{\mu}a^{\nu}
                  +{i\over g}[w_{\mu},w_{\nu}]D^{\mu}w^{\nu}
                  +{1\over 4g^2}[w_{\mu},w_{\nu}]^2
\bigr]\cr
&\hskip 55pt 1\hskip 45pt N^3\hskip 50pt N^2\hskip 65pt N^4\hskip 65pt
N^5\cr
%}}
%
%
%\eqn\SfN{\eqalign{
S_{\rm f}=&
\int\,\Tr\bigl[{i\over 2}\theta^T\gamma_{\mu}\partial^{\mu}\theta
                       +{i\over 2}\eta^T\gamma_{\mu}D^{\mu}\eta
                      +{1\over g}\eta^T\gamma_{\mu}[w^{\mu},\theta]
                     +{1\over 2g}\eta^T\gamma_{\mu}[w^{\mu},\eta]
\bigr]\cr
&\hskip 55pt 1\hskip 45pt N^2\hskip 50pt N^2\hskip 65pt N^3}}
and the ghost field contribution is
\eqn\SghN{\eqalign{
S_{\rm gh}=&
\int\,\Tr\bigl[\zeta^*(\partial_{\mu})^2 \zeta
                        +\omega^*(D_{\mu})^2\omega
                      -{i\over g}\omega^*[D_{\mu}w^{\mu},\zeta]
                      \cr
&\hskip 45pt 1\hskip 45pt N^3\hskip 55pt N^3\cr
&\hskip 100pt -{i\over g}\zeta^*\partial_{\mu}[w^{\mu},\omega]
                      -{i\over g}\omega^*D_{\mu}[w^{\mu},\omega]
              +{1\over g^2}\omega^*[w_{\mu},[w^{\mu},\omega]_{_{\rm diag}}]
\bigr]\cr
&\hskip 145pt N^2\hskip 50pt N^4\hskip 85pt N^4}}
The fields $\zeta$ and $\omega$ are respectively the diagonal and off
diagonal ghost fields.  Below each term is its corresponding factor of
$N$ coming from replacing the sums over indices by integrals and
rescaling the coordinates $(\tau,\sigma)$ to be independant of
$N$. The factors of $N$ associated with the covariant derivatives
$D_{\mu}=\partial_{\mu}-i/g[a_{\mu},\,\,\,\,]$ come from the commutator
with $a_{\mu}$. Rescaling the fields by 
\eqn\scfl{
w^{\mu}\rightarrow{1\over N^{3/2}}w^{\mu},\quad
\eta\rightarrow{1\over N}\eta,\quad
\omega\rightarrow{1\over N^{3/2}}\omega,}
we arrive at equations \Sb , \Strexp\ and \Sgsgo . There is one
subtlety : after rescaling the third term in the ghost action \SghN\
is of order $N^0$. It does not contribute at order $N^0$ however since
it must always be accompanied by the 4th term which (after rescaling)
is of order $1/N$ and which is the only term with the fields $\zeta^*$
and $\omega$.

\listrefs

\bye